\documentclass[12pt]{article}
\usepackage{latexsym}
\newcommand{\dd}{\mbox{\rm d}}
\newcommand{\DD}{\mbox{\rm D}}

\newcommand{\oo}{\over}
\newcommand{\p}{\partial}
\newcommand{\ci}{\cite}

\newcommand{\be}{\begin{equation}}
\newcommand{\ee}{\end{equation}}
\newcommand{\lbl}{\label}
\newcommand{\bi}{\bibitem}

\newcommand{\vs}{\vspace}
\newcommand{\hs}{\hspace}

\newcommand{\ggg}{\sqrt{|g(x')|}}

\def\bear{\begin{eqnarray}}
\def\ear{\end{eqnarray}}

\begin{document}
\baselineskip .7cm

\thispagestyle{empty}

\ 

\vs{1.5cm}

\begin{center}

{\bf \LARGE How the Geometric Calculus Resolves the Ordering Ambiguity
of Quantum Theory in Curved space}

\vs{6mm}

Matej Pav\v si\v c\footnote{Email: MATEJ.PAVSIC@IJS.SI}

Jo\v zef Stefan Institute, Jamova 39, SI-1000 Ljubljana, Slovenia

\vs{1.5cm}

ABSTRACT
\end{center}

The long standing problem of the ordering ambiguity in the definition of the
Hamilton operator for a point particle in curved space
is naturally resolved by using the powerful geometric calculus based on
Clifford Algebra. The momentum operator is defined to be the vector
derivative (the gradient) multiplied by $-i$; it can be expanded in terms
of basis vectors $\gamma_\mu$ as $p = -i \gamma^\mu \p_\mu$. The product of
two such operators is unambiguous, and such is the Hamiltonian which is
just the D'Alambert operator in curved space; the curvature scalar
term is not present in the Hamiltonian if we confine our consideration
to scalar wave functions only. It is also shown that $p$ is Hermitian
and self-adjoint operator: the presence of the basis vectors
$\gamma^\mu$ compensates the presence of $\sqrt{|g|}$ in the matrix elements
and in the scalar product. The expectation value of such operator follows
the classical geodetic line. 

\newpage

\section{Introduction}

An important step towards quantum gravity is formulation of quantum 
field theory in curved spacetime \ci{Birrell}. A commonly used method is DeWitt's
\ci{DeWitt1}--\ci{DeWitt3} curved space generalization of
the Fock-Schwinger proper time
technique \ci{Schwinger}
for finding the Feynman Green's function. In this connection, as a toy model
many researchers \ci{DeWitt1}--\ci{Grosche}
have studied the quantization of a non relativistic particle
in curved space $V_n$, whose dimension and signature can be left
unspecified.

The most straightforward approach has been in using the coordinate-based
formulation of the quantities such as the momentum $p_\mu$ and the Hamiltonian
which is proportional to the quadratic form 
$g^{\mu \nu} p_\mu p_\nu$, $\mu , \, \nu =1,...,n$.
When $p_\mu$ are quantum operators
the position of the curved space metric $g^{\mu \nu}$ (which depends on $x^\mu$)
does matter: the expression
$p_\mu g^{\mu \nu} p_\nu$ is different from $g^{\mu \nu} p_\mu p_\nu$ or
$p_\mu p_\nu g^{\mu \nu}$. This is the well known ordering ambiguity.
Different ordering prescriptions lead to different Hamilton operators which
in general contain a term with the scalar curvature $\alpha \hbar^2 R$.
Different authors, using different procedures, have obtained different values
for the coefficient $\alpha$. A related
problem is the correct definition of momentum operator $p_\mu$ which should
be Hermitian 
and retain the desirable properties under general coordinate transformations
of coordinates $x^\mu$. A resolution of the latter problem has been proposed
by DeWitt \ci{DeWitt1,DeWitt2}. But the ordering ambiguity has 
remained unresolved within the coordinate based operator formalism.

A promising more general line of research is geometric quantization 
\ci{Woodhouse} which employs coordinate-free differential geometry. If
applied to curved spacetime
there is no difficulties in defining a self-adjoint momentum operator
and no ordering problem does arise.

In this paper we explore the quantization in curved spacetime by using
yet another, also very powerful, geometric calculus which is based
on Clifford algebra \ci{Hestenes}. We define momentum as a vector 
$p= \gamma^\mu \p_\mu$ which is expanded in terms of the Clifford numbers
$\gamma^\mu$ which serve the role of basis vectors.
The presence of the basis vectors $\gamma^\mu$
besides the components $p_\mu$ renders the corresponding quantum operator
and its powers unambiguously defined and Hermitian. There is no ordering
ambiguity in this approach. We find that after quantization the Hamilton
operator (entering the Schr\" odinger equation for {\it scalar} wave
function) contains no term with the curvature scalar $R$. This result
agrees with that by Kleinert \ci{Kleinert} 
who has carefully examined the paths integral
quantization in curved spaces in a way which is distinct from other approaches
found in the literature \ci{DeWitt2,PathIntegral}.

The quantization based on geometric calculus opens an interesting possibility to
define the expectation $\langle p \rangle$ for momentum operator.
We show that $\langle p \rangle$ is real and that it is tangent to a
geodetic trajectory in curved space.

In Sec. 2 we briefly review the geometric calculus based on Clifford algebra
and the concept of {\it vector derivative}. In next section we provide a
definition of {\it vector integral}, an operation which is used
in Sec. 5 where we discuss the expectation value of the momentum operator.
In Sec.4 we consider quantization in curved space and write the matrix elements of
the momentum vector operator $p=\gamma^\mu p_\mu$ between position
eigenstates. We find that the latter matrix elements satisfy the condition
for Hermiticity without adding an extra term which is necessary
in the coordinate-based approaches \ci{DeWitt1}--\ci{Grosche}. We also show that
$\langle x| p^2 |\phi \rangle$ can be calculated by inserting twice the
complete set of the position eigenstates $|x \rangle$ and that the result is 
the covariant D'Alambert operator (multiplied by $-1$) as it should be.
So we have succeeded in providing a consistent and elegant foundations for
quantum mechanics in curved space.

\section{Geometric calculus}

Since the seminal work by Hestenes \ci{Hestenes}, together with some very 
important
works such as those of refs. \ci{Clifford}, the geometric calculus based on
Clifford algebra has been receiving increasing attention \ci{Clifford1}. 
It is a
very powerful language for geometry and physics, enabling potentially
very important generalizations \ci{Pezzaglia, Castro,Pavsic1,Castro-Pavsic}.
The calculus is based on the observation that the non commuting numbers
satisfying the Clifford algebra relations
\be
     \gamma_\mu \cdot \gamma_\nu \equiv \mbox{$1\oo 2$} (\gamma_\mu \gamma_\nu +
     \gamma_\nu \gamma_\mu) = g_{\mu \nu}
\lbl{1}
\ee
can represent basis vectors. An arbitrary vector $a$ is then a linear
superposition
\be
    a= a^\mu \gamma_\mu
\lbl{2}
\ee
where the components $a^\mu$ are {\it scalars} from the geometric
point of view, whilst $\gamma_\mu$ are {\it vectors}.
The latter relation is just a more general form of a similar relation
discussed in textbooks:
\be
        a = a_0 {\underline 1} + a^i \sigma_i
\lbl{3}
\ee
which states that a 4-vector can be written in terms of the Pauli matrices
$\sigma_i$. Here we do not consider $\gamma_\mu$ as being necessarily
matrices\footnote{Also the usual imaginary number $i$, satisfying $i^2 = -1$,
is not considered as a matrix, although it can be represented as a matrix
of real numbers.};
they are just numbers satisfying eq.(\ref{1}).

Besides the basis $\gamma_\mu$ we can introduce the dual basis $\gamma^\mu$
satisfying
\be
      \gamma^\mu \cdot \gamma^\nu \equiv \mbox{${1\oo 2}$} 
      (\gamma^\mu \gamma^\nu + \gamma ^\nu \gamma^\mu  )= g^{\mu \nu}
\lbl{4}
\ee
where $g^{\mu \nu}$ is the covariant metric tensor such that $g^{\mu \alpha}
g_{\alpha \nu} = {\delta^\mu}_\nu$, $\; \gamma^\mu \gamma_\nu + \gamma_\nu
\gamma^\mu = 2 {\delta^\mu}_\nu$ and $\gamma^\mu = g^{\mu \nu} \gamma_\nu$.
We shall consider {\it curved space} in which $\gamma_\mu$ and $g_{\mu \nu}$
depend on position $x$.

The {\it vector derivative} or {\it gradient} is defined according to
\be
      \p \equiv \gamma^\mu \p_\mu
\lbl{5}
\ee
where $\p_\mu$ is an operator, that I will call {\it geometric derivative},
whose action depends on the quantity it acts on.

Applying the vector derivative $\p$ on a {\it scalar} $\phi$ we have
\be
    \p \phi = \gamma^\mu \p_\mu \phi
\lbl{5a}
\ee
where $\p_\mu \phi \equiv (\p  /\p x^\mu )\phi$ coincides with the partial 
derivative of $\phi$.

But if we apply it on a {\it vector} $a$ we have
\be
     \p a = \gamma^\mu \p_\mu (a^\nu \gamma_\nu) = \gamma^\mu (\p_\mu a^\nu
     \gamma_\nu + a^\nu \p_\mu \gamma_\nu)
\lbl{6}
\ee
In general $\gamma_\nu$ is not constant; it satisfies the relation
\ci{Hestenes, Pavsic1}
\be
     \p_\mu \gamma_\nu = \Gamma_{\mu \nu}^\alpha \gamma_\alpha
\lbl{7}
\ee
where $\Gamma_{\mu \nu}^\alpha$ is the {\it connection}. Similarly, for
$\gamma^\nu = g^{\nu \alpha} \gamma_\alpha$ we have
\be
     \p_\mu \gamma^\nu = - \Gamma_{\mu \alpha}^\nu \gamma^\alpha
\lbl{7a}
\ee
The {\it non commuting} operator $\p_\mu$ so defined determines
{\it the parallel
transport} of a basis vector $\gamma^\nu$. It should be distinguished
from the ordinary---{\it commuting}---partial derivative 
${\gamma^\nu}_{,\mu}$ as defined in eq.(\ref{A5}).
 Instead of the symbol $\p_\mu$
Hestenes uses $\Box_\mu$, whilst Wheeler et. al. \ci{Wheeler} use
$\nabla_\mu$ and call it ``covariant derivative". We find it convenient
to keep the symbol $\p_\mu$ for components of the geometric
operator $\p = \gamma^\mu \p_\mu$. When acting on a scalar field
the geometric derivative $\p_\mu$ happens to be commuting and thus behaves
as the ordinary partial derivative.

Using (\ref{7}), eq.(\ref{6}) becomes  
     $$\p a = \gamma^\mu \gamma_\nu (\p_\mu a^\nu + \Gamma_{\mu \alpha}^\nu
     a^\alpha) \equiv \gamma^\mu \gamma_\nu \DD_\mu a^\nu = \gamma^\mu
     \gamma^\nu \DD_\mu a_\nu$$
\be
     = \gamma^\mu \cdot \gamma^\nu \DD_\mu a_\nu + \gamma^\mu \wedge
     \gamma^\nu \DD_\mu a_\nu     
\lbl{8}
\ee
where $\gamma^\mu \cdot \gamma^\nu$ is the {\it symmetric} and
$\gamma^\mu \wedge \gamma^\nu \equiv {1\oo 2} (\gamma^\mu \gamma^\nu -
\gamma^\nu \gamma^\mu)$ is the antisymmetric part of the
{\it Clifford product} $\gamma^\mu \gamma^\nu$. In general we have
\be
     \p^n a = \gamma^{\mu_1} \gamma^{\mu_2} ... \gamma^{\mu_n} \gamma^\nu
     \DD_{\mu_1} \DD_{\mu_2} ... \DD_{\mu_n} a_\nu
\lbl{8a}
\ee     
Here $\DD_\mu$ is the {\it covariant derivative} of the tensor calculus
with the properties:
\bear
       \DD_\mu \phi &=& \p_\mu \phi \nonumber \\
       \DD_\mu A^\nu &=& \p_\mu A^\nu + \Gamma_{\mu \rho}^\nu A^\rho \nonumber \\
       \mbox{etc .}&&
\lbl{8b}
\ear       

From the relation (\ref{8a}) we find that after applying the vector derivative
on a scalar twice we obtain
\bear
    \p \p \phi &=& (\gamma^\mu \p_\mu)(\gamma^\nu \p_\nu) \phi
   = \gamma^\mu \gamma^\nu \DD_\mu \DD_\nu \phi \nonumber \\
     &=& \gamma^\mu \cdot \gamma^\nu \DD_\mu \DD_\nu \phi + \mbox{$1\oo 2$}
     \gamma^\mu \wedge \gamma^\nu (\DD_\mu \DD_\nu - \DD_\nu \DD_\mu) \phi
\lbl{9}
\ear
If the connection is symmetric, i.e., when torsion $C_{\mu \nu}^\rho
= \Gamma_{\mu \nu}^\rho - \Gamma_{\nu \mu}^\rho$ is zero, the commutator
of the covariant derivatives acting on a {\it scalar} vanishes, 
$[\DD_\mu , \DD_\nu] \phi = - C_{\mu \nu}^\rho = 0$, and we have
\be
     \p \p \phi = g^{\mu \nu} \DD_\mu \DD_\nu \phi = \DD_\mu \DD^\mu \phi
     = {1\oo {\sqrt{|g|}}} \p_\mu (\sqrt{|g|} \, g^{\mu \nu} \p_\nu \phi )
\lbl{9a}
\ee
which is just the D'Alambert operator in curved space. Here, as usually,
$g \equiv {\rm det}\, g_{\mu \nu}$ denotes the determinant of the metric.

A more elaborated discussion of the geometric calculus the reader can find
in refs. \ci{Hestenes,Pavsic1,Castro-Pavsic}. Here let me just mention
that in geometric calculus the sum of scalars, vectors $a^\mu \gamma_\mu$,
bivectors $a^{\mu \nu} \gamma_\mu \wedge \gamma_\nu$, etc., is as legitimate
operation as is the sum of a real and imaginary number.

This formalism should not be confused with a different formalism discussed
by \ci{Hehl} and \ci{Mielke}, where not only Clifford numbers, but also
differential forms take place.

The equations of geometric calculus considered above do not rely on
a particular representation of Clifford algebra and thus hold in space
of any dimension, including dimension 1. In the latter case the set of basis
vectors consists of only one element $\gamma_0$ whose square gives a single
component of the metric $g_{00} = \gamma_0 \cdot \gamma_0 = \gamma_0^2$
which in general depends on position. By a suitable coordinate transformation
it can be transformed to $g'_{00} = (\p x/\p x')^2 g_{00} = 1.$

In this paper we do not consider the action of the derivative operator
(\ref{5}) on a {\it spinor}. This could be done straightforwardly by taking
into account the well know fact that spinors can be represented 
\ci{Hestenes,Clifford,Pavsic1,Mankoc} as
members of the left or right ideals of Clifford algebra.

\section{Vector integral in geometric calculus}

Before going to physics, more precisely, to quantum mechanics in curved
space, some more mathematical preparations are necessary.
First let us recall that for an arbitrary vector with components
$a^\mu$ the expression
\be
    {\bar a}^\mu = \int_\Omega \dd^n x \, \sqrt{|g|} \,  a^\mu  
\lbl{15}
\ee
has no geometric sense, since it depends on a chosen coordinate system and
has no definite
properties under general coordinates transformations.

On the other hand, if we integrate not {\it components} $a^\mu$ but a 
{\it vector}
$a = a^\mu \gamma_\mu$ we obtain an expression which is invariant under
arbitrary coordinates transformations:
\be
    {\bar a} = \int_\Omega \dd^n x \, \sqrt{|g|} \,  a^\mu \gamma_\mu 
\lbl{16}
\ee
This is so, because a vector $a = a^\mu \gamma_\mu$ by definition
does not depend on coordinates. Under a change of coordinates the transformation
of components
\be
     a'^\mu = {{\p x'^\mu}\oo {\p x^\alpha}} a^\alpha
\lbl{17}
\ee
is compensated by the corresponding transformation of basis vectors
\be
     \gamma'_\mu = {{\p x^\beta}\oo {\p x'^\mu}} \gamma_\beta
\lbl{18a}
\ee
so that the vector $a = a^\mu \gamma_\mu = a'^\mu \gamma'_\mu$ remains
invariant. Since the volume element $\sqrt{|g|} \, \dd^n x$ also is invariant, 
the ``average" ${\bar a}$ in eq.(\ref{16}) is an invariant quantity.

A question arises at this point of whether the integration (\ref{16})
over a vector field is a well defined operation from the geometric point
of view: the
result should be a geometric object---a tangent vector---of the considered curved
space. We shall show that this can indeed be the case.
Namely, if we have some rule for bringing together vectors at
different points and integrate them at a chosen point of the manifold, then
such operation is geometrically well defined. We will see that the integral
(\ref{16}), if properly interpreted, can in fact incorporate such a
rule.

Let us now examine the integral (\ref{16}) more closely. 
First let us introduce at every 
point $x$ a local ``Lorentz" frame\footnote{
If we interprete our formalism as describing the non relativistic theory in
an $n$-dimensional curved Euclidean space, then $\gamma^a$ at every point
$x$ span a local frame of a flat space tangent to $x$. Then instead of a local
Lorentz transformation we have a local rotation.}
spanned by a set of basis vectors $\gamma_a$
and their duals $\gamma^a = \eta^{ab} \gamma_b$ satisfying
\be
     \gamma_a \cdot \gamma_b = \eta_{ab} \; , \qquad 
     \gamma^a \cdot \gamma^b = \eta^{ab}
\lbl{A1}
\ee
where $\eta_{ab}$ is the Minkowski metric and $\eta^{ab}$ its inverse.
At every point  the basis vectors
$\gamma_\mu$ can be expressed in terms of $\gamma_a$, and vice versa:
\be
      \gamma_\mu = (\gamma_\mu \cdot \gamma^a) \gamma_a \: , \qquad
      \gamma_a = (\gamma_a \cdot \gamma^\mu) \gamma_\mu 
\lbl{A2}
\ee
where
\be
    \gamma_\mu \cdot \gamma^a = {e^a}_\mu \; , \qquad 
    \gamma_a \cdot \gamma^\mu = {e_a}^\mu
\lbl{A3}
\ee
are the {\it vielbein field} and its inverse, respectively. Inserting
({\ref{A2}) into the integral (\ref{16}) we obtain
\be
     {\bar a} = \int_\Omega \dd^n x \, \sqrt{|g|} \, a^\mu (x) {e^a}_\mu (x) 
     \gamma_a (x)
\lbl{A4}
\ee

The above expression (\ref{A4}) is invariant with respect to arbitrary general
coordinate transformations of spacetime coordinates $x^\mu$ (passive
diffeomorphisms). In addition, it is invariant with respect to arbitrary
{\it passive} local Lorentz transformations (which act on the index $a$),
since any transformation of ${e^a}_\mu$ is compensated by the corresponding
transformation of $\gamma_a$ so that that the integrand remains the
same vector at $x$. Consequently, the integral itself remains invariant.

Choice of a local
Lorentz frame at every point $x$ is in fact a choice of gauge, namely
a local Lorentz gauge. The result of the integration (\ref{A4})
does not depend on choice of gauge. Therefore, in order to perform
calculation, we are free to take whatever gauge we find convenient. This
will have no influence on our result. If $\gamma_a$ are represented as matrices
satisfying (\ref{A1}) the result of the integration is a matrix ${\bar a}$,
and the matrix is the same regardless of which local Lorentz gauge we choose
in the integral (\ref{A4}). Does the matrix ${\bar a}$ represents a vector of
our manifold? In other words, can ${\bar a}$ be expanded in terms of basis
vectors $\gamma_a$ or $\gamma_\mu$ taken at a chosen point $x'$ within a domain
$\Omega$ of the manifold over which the integration is performed?

Let us choose a gauge in which $\gamma_a$ are constant\footnote{
Constructing a curved space expression by means of constant flat space matrices
$\gamma_a$ and vielbein ${e^a}_\mu$ is a standard procedure (c.f. the action
for the Dirac particle in curved spacetime).}
(see eqs. (\ref{A5})--(\ref{A7})) at every point $x$.
Then they can be taken out of the integral and so (\ref{A4}) becomes
\be
    {\bar a} = \left ( \int_{\Omega} \dd^n x \, \sqrt{|g|} \, a^\mu (x) 
    {e^a}_\mu (x) \right ) \gamma_a \equiv A^a \gamma_a
\lbl{A11}    
\ee
The result of the integration (\ref{A11}) over a chosen domain $\Omega$
in the manifold
is the object ${\bar a} \equiv A^a \gamma_a$. If the Lorentz basis vectors 
$\gamma_a$ are represented 
as Dirac {\it matrices} in flat space, then also ${\bar a}$ is a matrix.
Since it can be expanded in terms of a complete set of basis vectors, ${\bar a}$
itself is a {\it vector}.

Although the integral (\ref{A11}) is invariant with respect to {\it passive}
local Lorentz transformations which change ${e^a}_\mu$ and $\gamma_a$, there is
still a freedom to perform at every point $x$ an {\it active} Lorentz
transformation
\be
      {e^a}_\mu (x) \rightarrow {e'^a}_\mu (x) = {L^a}_b (x) {e^b}_\mu (x)
\lbl{B1}
\ee
whilst keeping $\gamma_a$ fixed. Under the transformations (\ref{B1}) the
metric does not change:
\be
       \gamma'_\mu \cdot \gamma'_\nu = \gamma_\mu \cdot \gamma_\nu = g_{\mu \nu}
\lbl{B2}
\ee

Here $\gamma_\mu = {e^a}_\mu \gamma_a$ and 
$\gamma'_\mu = {e'^a}_\mu \gamma_a$ are
two different Clifford numbers that can be represented by two different
matrices which solve the Clifford algebra relation (\ref{B2}). The integral
(\ref{16}) or its variants (\ref{A4}), (\ref{A11}) depend on choice of
$\gamma_\mu (x)$, i.e., on choice of ${e^a}_\mu (x)$ and $\gamma_a (x)$.
Since the choice of $\gamma_\mu (x)$ is arbitrary, it seems at first sight
that there is no way of defining the integral (\ref{16}) uniquely. But is we
think deeper, we observe that in flat space the integral, if written in the
general form (\ref{A4}), is not uniquely defined as well. In flat space we
also can distinguish between the coordinate frame $\gamma_\mu$, satisfying
$\gamma_\mu \cdot \gamma_\nu = g_{\mu \nu} (x)$, and the local Lorentz frame
$\gamma_a$, satisfying $\gamma_a \cdot \gamma_b = \eta_{ab}$. There is still
a lot of freedom in choosing a representation for $\gamma_\mu$ and $\gamma_a$
at every point $x$. But there is a choice which distinguishes itself
from other possible choices. In flat space such a choice is straightforward.
My proposal is now that when considering the integral in curved space,
choice of $\gamma_\mu$ and $\gamma_a$ has to be such that in the flat space
limit eq. (\ref{A4}) gives the well known result for the integration in flat
space. In the following I will discuss such a choice.

Let us consider a curved space $V_n$ and let us choose a point $x'$ and
the quantities ${e^a}_\mu (x')$, $\gamma_a (x')$, $\gamma_\mu (x') = 
{e^a}_\mu (x') 
\gamma_a (x')$ at that point. Then at every other point $x \in \Omega$
we choose $n$ vectors $\gamma^a (x) = {e^a}_\mu (x) \gamma^\mu (x)$,
$a=1,2,...,n$, such that
they will be parallel to the vectors $\gamma^a (x') = {e^a}_\mu (x') 
\gamma^\mu (x')$ in the sense of the parallel transport along 
the geodesic\footnote{We assume here that for any two given points within the
domain $\Omega$ of integration there exists a geodesic, and that it is
unique. The cases where this is not true, are excluded from our consideration.
We assume that in a curved space which allows for non existence or non uniqueness
of geodesics, the integration domain has to be narrowed down to such extent,
that the above complication no longer occurs. We do not claim that our
integral can be defined globally.}
joining the point $x$ and $x'$. Under the parallel transport the components
${e^a}_\mu$ of $\gamma^a$ transform according to
\be
    {e^a}_\mu (x) = {g_\mu}^\nu (x,x') {e^a}_\nu (x')
\lbl{B3}
\ee
where ${g_\mu}^\nu (x,x')$ is the {\it parallel propagator} \ci{Synge}. Inverting
(\ref{B3}) we find
\be
     {g_\mu}^\nu (x,x') = {e^a}_\mu (x) {e_a}^\nu (x')
\lbl{B3a}
\ee
Further, we represent $\gamma^a (x')$ and $\gamma^a (x)$ by the same
flat space matrices so that
\be
       \gamma^a (x) = \gamma^a (x')
\lbl{B4}
\ee
From eq. (\ref{B4}) we have
\be
     {e^a}_\mu (x) \gamma^\mu (x) = {e^a}_\mu (x') \gamma^\mu (x')
\lbl{B5}
\ee
which gives
\be
     \gamma^\nu (x) = {e_a}^\nu (x) {e^a}_\mu (x') \gamma^\mu (x')
     = {g^\nu}_\mu (x, x') \gamma^\mu (x')
\lbl{B6}
\ee
The latter restriction on $\gamma^\nu (x)$ is a consequence of choice
(\ref{B4}).

Inserting (\ref{B3}) and (\ref{B4}) into the integral (\ref{A4}) we find
\bear
    {\bar a} &=& \int \dd^n x \, \sqrt{|g|} \, a^\mu (x) {g_\mu}^\nu (x,x')
     {e^a}_\nu (x') \gamma_a (x') \nonumber \\
     &=& \int \dd^n x \, \sqrt{|g|} \, a^\mu (x) {g_\mu}^\nu (x,x') \gamma_\nu
     (x') \nonumber \\
     &=& \int \dd^n x \, \sqrt{|g|} \, a^\nu (x',x) \gamma_\nu (x')
     \equiv A^\nu (x') \gamma_\nu (x')
\lbl{B7}
\ear
where
\be
     a^\nu (x', x) = a^\mu (x) {g_\mu}^\nu (x,x') 
\lbl{B8}
\ee  
This is the relation for {\it parallel transport} of a vector with components
$a^\mu (x)$ from a point $x$ along the geodesic to a point $x'$ (see, e.g.,
\ci{Synge}).

In eq.(\ref{B7}) vectors are first parallelly transported from points 
$x \in \Omega$ to a chosen point $x'$ and then they are integrated. This
obvious result, that also holds in flat space, has been obtained from the
general expression (\ref{A4}) for the choice of ${e^a}_\mu$ as given in
eq. (\ref{B3}) and following the procedure of eqs. (\ref{B4})--(\ref{B7}).

Clearly $A^\nu (x') \gamma_\nu (x')$ in eq.(\ref{B7}) is the vector at 
$x'$ that is obtained by summing
(more precisely, integrating) the vectors
\be
         a^\nu (x',x) \gamma_\nu (x') = a^\mu (x) {e^a}_\mu (x) {e_a}^\nu (x') 
         \gamma_\nu (x')
\lbl{A12d}
\ee
which are all taken at the same point $x'$ (which can be any point
within the domain $\Omega$). The argument $x$ of 
the integration determines the points from which the vectors
$a^\nu (x',x)$ were brought by means of the parallel transport along the
geodesic joining $x$ and $x'$. The integral (\ref{16}) was thus shown to
have a well defined geometrical meaning. {\it The integral by its construction
and the choices (\ref{B3}),(\ref{B4}) automatically implies the rule for
bringing together vectors at different points.} So
far it has been taken for granted that in the integral such as (\ref{16})
we are summing vectors at different points and therefore such an operation
has no geometric sense.
It has not been realized that if in the definition of the integral (\ref{16})
we employ the choices (\ref{B3}),(\ref{B4}), then vectors are actually
integrated at the same point of the manifold.
In curved space we are doing exactly the same as in flat
space: we first bring vectors together by parallel transport and then
perform the summation or the integration. Eq. (\ref{16}) is nothing but
a short notation for such operation.

We have shown that the integral (\ref{16}) can be
written in terms of $\gamma_a$ that are ``constant" at every point $x$.
In order to understand what a constant $\gamma_a$ precisely means, let us
recall that we distinguish between two different types of derivative:

\ (i) {\it Geometric derivative} (determining the parallel transport)
gives
\bear
        &&\p_\nu \gamma_\mu = \Gamma_{\nu \mu}^\rho \gamma_\rho \lbl{A8} \\
        &&\p_\nu \gamma_a = - {{\omega_a}^b}_\nu \gamma_b \lbl{A9} \\
        &&\p_\nu {e^a}_\mu = \Gamma_{\nu \mu}^\rho {e^a}_\rho - 
        {\omega^{ab}}_\nu  e_{b \mu} \lbl{A7a}
\ear
where ${\omega^{ab}}_\mu$ is the connection for the local Lorentz frame
field $\gamma_a$. 

(ii) The commuting {\it partial derivative} satisfies
\bear
    &&\gamma_{\mu , \nu} = \Gamma_{\nu \mu}^\rho \gamma_\rho
    - \omega_{ab \nu} {e^b}_\mu \gamma^a + {e^a}_\mu \gamma_{a,\nu} \lbl{A5} \\
    &&\gamma_{a,\nu} = \mbox{arbitrary (in general may be different from zero)} 
    \lbl{A6} \\
    &&{e^a}_{\mu ,\nu} = \Gamma_{\nu \mu}^\rho {e^a}_\rho - {\omega^{ab}}_\nu 
    e_{b \mu} \lbl{A7}
\ear
Since  ${e^a}_\mu$ are the {\it scalar} components of
the vector $\gamma^a = {e^a}_\mu \gamma^\mu$, the geometric and the
partial derivative of ${e^a}_\mu$ coincide.

Eqs. (\ref{A8})--(\ref{A7}) hold for arbitrary $\gamma_a$ and they 
are covariant with respect to general coordinate
transformations (\ref{17}),(\ref{18a}) and local Lorentz transformations
$\gamma'_a = {L_a}^b \gamma_b$ provided that the connections transform as
\bear
      {\Gamma'}_{\nu \mu}^\rho &=& {{\p x'^\rho}\oo {\p x^\sigma}}
      {{\p x^\alpha}\oo {\p x'^\mu}} {{\p x^\beta}\oo {\p x'^\nu}} \, 
      \Gamma_{\alpha \beta}^\sigma + {{\p x'^\rho}\oo {\p x^\sigma}}
      {{\p^2 x^\sigma}\oo {\p x'^\mu \p x'^\nu}} \lbl{con1} \\
      \nonumber \\
      {\omega'^{ab}}_\mu &=& {\omega^{cd}}_\mu {L^a}_c {L^b}_d +
      {L^a}_c {L^{cb}}_{,\mu} \lbl{con2}
\ear

If $\gamma_{a,\nu} = 0$, $a=1,2,...,n$, $\nu=1,2,...,n$, i.e., if the 
(commuting) partial derivative of $\gamma_a$ is zero, then we say
that $\gamma_a$ is constant at every point $x$ of the manifold.

Taking the {\it geometric derivative} of an arbitrary vector $A$
we have after using eq.(\ref{A9}):
\be
     \p_\nu A = \p_\nu (A^a \gamma_a) = (\p_\nu A^a + {\omega^a}_{b \nu} A^b)
     \gamma_a \equiv {\DD_\mu} A^a \, \gamma_a
\lbl{A15}
\ee
On the other hand, using eq.(\ref{A8}), we have
\be
    \p_\nu A = \p_\nu (A^\mu \gamma_\mu) = (\p_\nu A^\mu  +
    \Gamma_{\nu \rho}^\mu \, A^\rho) \gamma_\mu \equiv \DD_\nu A^\mu \gamma_\mu
\lbl{A16}
\ee
where $\DD_\nu A^\mu \equiv {A^\mu}_{;\nu} = \p_\nu A^\mu + 
\Gamma_{\nu \rho}^\mu A^\rho$
is just the covariant derivative of a vector field. Comparing (\ref{A15}) 
and (\ref{A16}) we find that
\be
    \DD_\nu A^\mu \gamma_\mu = \DD_\mu A^a \, \gamma_a
\lbl{A17}
\ee

Taking the commutators of the geometric derivatives acting on 
$A = A^\mu \gamma_\mu$ we obtain
\be
      [\p_\alpha , \p_\beta] A = A^\mu [\p_\alpha , \p_\beta] \gamma_\mu =
      ([\DD_\alpha , \DD_\beta ] A^\mu) \gamma_\mu = {R^\mu}_{\nu \alpha \beta}
      \, A^\nu \, \gamma_\mu
\lbl{A18}
\ee
where
\be
    {R^\mu}_{\nu \alpha \beta} = \p_\beta \Gamma_{\nu \alpha}^\mu -
    \p_\alpha \Gamma_{\nu \beta}^\mu + \Gamma_{\beta \rho}^\mu \, 
    \Gamma_{\alpha \nu}^\rho - \Gamma_{\alpha \rho}^\mu \, 
    \Gamma_{\beta \nu}^\rho
\lbl{A18a}
\ee
If acting on $A= A^a \gamma_a$, the commutator gives
\be
   [\p_\alpha , \p_\beta] A = A^a [\p_\alpha , \p_\beta] \gamma_a  =
   ([\DD_\alpha , \DD_\beta ] A^a) \gamma_a  = R_{ab \alpha \beta} 
   \, A^b \gamma^a
\lbl{A18b}
\ee
where
\be
      R_{ab \alpha \beta} = - (\omega_{ab \alpha ,\beta} - \omega_{ab
   \beta , \alpha} + \omega_{ac \alpha} {\omega^c}_{b \beta} -
   \omega_{ac \beta} {\omega^c}_{b \alpha} )
\lbl{A18c}
\ee

If, in particular, we take for A just a vector $\gamma_a = {e_a}^\mu \gamma_\mu$,
$a = 1,2,...,n$, we find from (\ref{A15}) that
\be      
   {{e_a}^\mu}_{;\nu} = {{e_a}^\mu}_{,\nu} + \Gamma_{\nu \rho}^\mu {e_a}^\rho =
   - {{\omega_a}^b}_\nu {e_b}^\nu
\lbl{A19}
\ee
which is consistent with eq.(\ref{A7a}) or  (\ref{A7}) for the derivative of
vielbein. So we have verified that the relation (\ref{B4}) for `` constant"
$\gamma_a$ is consistent with eqs. (\ref{A8})--(\ref{A7a}) for geometric
derivative. If in a curved manifold the field $\gamma_a (x)$ is constant
in the sense that $\gamma_{a, \mu} = 0$, then this reflects merely a choice
of the representation for $\gamma_a$ (a solution to the Clifford algebra
relation (\ref{A1})) at every point $x$. We have shown that when expanded
in terms of the basis vectors $\gamma_\mu (x)$ according to $\gamma_a =
{e_a}^\mu \, \gamma_\mu$ the vector field $\gamma_a$ behaves correctly
from the geometric point of view. If vectors $\gamma_a (x)$, $a=1,2,...,n$ for
any $x \in \Omega$ are parallel (in the sense of parallel transport
along a geodesic) to corresponding vectors $\gamma_a (x')$ at a given
point $x'$, this does not imply that vectors at different points
$x \neq x'$ are parallel amongst themselves.

\paragraph{Extrinsic and intrinsic integral} In our approach only the intrinsic
geometry of the manifold has been considered
and the integral has been given the geometric meaning which is intrinsic
to the manifold. Had we considered our manifold $V_n$ as being embedded in
a higher dimensional space $V_N$, then we would have two options of how
to perform (or better, to define) the integration:

 \ (i) either intrinsically, by using the intrinsic basis vectors $\gamma_\mu$ 
and $\gamma_a$ satisfying the relations
(\ref{A8})--(\ref{A9}) and choosing a gauge in which $\gamma_a$
are constant according to eq.(\ref{A6}),

 (ii) or extrinsically, by considering $\gamma_\mu$ and $\gamma_a$ as being
induced from the corresponding vectors in $V_N$ and taking the analog
of the relations (\ref{A8})--(\ref{A9}) in the embedding space $V_N$.

That is, when we write an expression such as (\ref{16}) we have to
specify (e.g., by using an extra label) over which manifold the integration
is to be performed. So we {\it define} two different integrals:
\be
     {\bar a} = \int \dd^n x \, \sqrt{|g|}\, a^\mu \gamma_\mu \biggl\vert_{V_n}
     = \int \dd^n x \, \sqrt{|g|}\, a^\mu {e^a}_\mu \gamma_a \biggl\vert_{V_n}
     = \mbox{vector in $V_n$}
\lbl{A20}
\ee
\bear
      &&{\bar A} = \int \dd^n x \, \sqrt{|g|}\, a^\mu \gamma_\mu \biggl\vert_{V_N} 
      = \int \dd^n x \, \sqrt{|g|}\, a^\mu \p_\mu \eta^M \gamma_M \biggl\vert_{V_N}
      \hs{4cm} \nonumber \\
      && \hs{4.5cm}= \int \dd^n x \, \sqrt{|g|}\, a^\mu \p_\mu \eta^M {E^A}_M
      \gamma_A
      \biggl\vert_{V_N} = \mbox{vector in $V_N$}
\lbl{A21}
\ear
Here $\eta^M (x^\mu)$, $M=1,2,...,N$, are the embedding functions for the
manifold $V_n$ which is considered as an $n$-dimensional surface embedded in
$V_N$. The quantities $\gamma_M$, $\gamma_A$ and ${E^A}_M = \gamma^A
\cdot \gamma_M$ they all refer to the embedding space $V_N$ and are
the basis vectors, the local Lorentz vectors and the vielbein, respectively.

In eq.(\ref{A20}),(\ref{A21}) $\gamma_\mu |_{_{_{V_n}}}$ and 
$\gamma_\mu  |_{_{_{V_N}}}$ are totally different objects: the former are
generators of the Clifford algebra of $V_n$, whilst the latter are the
linear combinations of the generators $\gamma_M$ of the Clifford algebra
of $V_N$. If represented as matrices, then $\gamma_\mu |_{_{_{V_n}}}$
and  $\gamma_\mu  |_{_{_{V_N}}}$ are completely different kinds of matrices,
although they both represent the tangent vectors to $V_n$.
This clarifies why the vector integral such as (\ref{16})
can be consistently defined either as a geometric object residing in 
the considered manifold $V_n$, or alternatively as a geometric object
residing in the embedding space $V_N$.

As an example imagine a curved 3-dimensional surface $V_3$ embedded in
flat 4-dimensional space(time) $V_4$. There are two possibilities:

  \ (i) Tangent vectors of $V_3$ can be
expanded in terms of three position dependent $2 \times 2$ matrices 
$\sigma_\alpha$, $\alpha = 1,2,3$, which in
turn can be expressed by means of the ``dreibein" in terms of $2 \times 2$
Pauli matrices $\sigma_i$, $i = 1,2,3$. Summation (or integration) of those
tangent vectors gives an object which is itself a tangent vector.

(ii) On the other hand, tangent vectors of $V_3$ can be induced from $V_4$, i.e.,
they can be considered from the extrinsic point of view and expanded
in term of the $4 \times 4$ matrices $\gamma_\mu$, $\mu = 0,1,2,3$.
It is clear now that in this (extrinsic) case the summation (or the
integration) of the tangent vectors gives a vector in $V_4$ which is
not necessarily a tangent vector of $V_3$.

Whilst in Case (ii) tangent vectors are $4 \times 4$ matrices that can
be expanded in terms of matrices $\gamma_\mu$, in Case (i) there 
is no way to express 
the tangent vectors---represented as $2 \times 2$ matrices---in terms
of $4 \times 4$ matrices $\gamma_\mu$. In the case (i) tangent vectors
are true intrinsic vectors to $V_3$ and they bear no relation to the
embedding space. 

{\it Summary - \ }We are performing quite a legitimate operation of
transferring vectors
$a(x)$ form $x \in \Omega$ into into a chosen point $x'$ and integrating
them at $x'$. As a result we obtain a vector ${\bar a} (x')$ at $x'$.

By means of geometric calculus based on Clifford algebra such operation
can be defined by employing the notation
\be
     {\bar a} = \int_{\Omega\{ V_n, x'\} } \dd^n x \, \sqrt{|g|} \, 
     a^\mu (x) \gamma_\mu (x)
\lbl{ A22}
\ee
where $\gamma_\mu (x) = {e^a}_\mu (x) \gamma_a (x)$ are expanded in terms of
vectors $\gamma_a (x)$ which are all chosen to be parallel to $\gamma_a (x')$
at $x'$. Mutually, of course, they are not parallel. For simplicity reasons
we then omit the subscript $\{ V_n , x' \}$ at the integration symbol.

Choice of the local Lorentz frame field $\gamma_a (x)$, or, equivalently,
${e^a}_\mu (x)$ is arbitrary. Choice must be such that the integral
coincides with the integral in flat space. Also in flat space it depends
on choice of $\gamma_a (x)$, that is, ${e^a}_\mu (x)$. We choose vectors
of the field $\gamma_a (x)$ all parallel to a given vector $\gamma_a (x')$,
and obtain the usual result for the vector integral in flat space. The
same choice of $\gamma_a (x)$ (that is, of ${e^a}_\mu (x)$) 
we keep in curved space.
If our space $V_n$ is embedded in a higher dimensional space $V_N$, it must
also be specified with respect to which space, $V_n$ or $V_N$, the
parallelism of $\gamma_a (x)$ and $\gamma_a (x')$ is taken. The parallelism
of $\gamma_a (x)$ and $\gamma_a (x')$ with respect to $V_n$ defines the
integral which is different from the integral in which the parallelism
of $\gamma_a (x)$ and $\gamma_a (x')$ is taken with respect to $V_N$.

To sum up, the vector integral (\ref{A4}) is a functional of the frame field
$\gamma_a (x)$ (or, equivalently, ${e^a}_\mu (x)$). In other words,
the integral ${\bar a}$ is an object which is defined with respect to a chosen
local Lorentz frame field. It is convenient to choose the frame field
such that in the transition from a curved to flat space the curved space
integral coincides with the usual flat space integral.

\section{Fock-Schwinger-DeWitt proper time approach to quantum theory
in curved space and geometric calculus}

When discussing the problem of quantum theory in a curved space $V_n$
the authors usually start from the following classical action
\be
     I[X^\mu] = {1\oo 2 \Lambda} \int \dd \tau \, g_{\mu \nu} {\dot X}^{\mu}
     {\dot X}^\nu
\lbl{C1}
\ee
where $\Lambda$ is a fixed constant and  $\tau$ an arbitrary parameter, whilst
$X^\mu$ , $\mu = 1,2,...,n$, are $\tau$-dependent functions denoting
position of the ``particle" in an $n$-dimensional space $V_n$.
Usage of the above action for description of a relativistic particle
has to be taken with some caution. The topics has been extensively discussed
in the literature. In relation to the usual relativistic theory
which starts from the well known minimal length action and in which momentum
is constrained to a mass shell,
there are at least three different possible
interpretations of the classical and quantized theory based on the
action (\ref{C1}). According to one interpretation (\ref{C1}) is a gauge
fixed action (i.e., an action in which reparametrization of $\tau$ is fixed). 
It is a specail case (for $p=0$) of the Schild action \ci{Schild}
which can be used
for description of $p$-branes. According to another interpretation,
(\ref{C1}) is an unconstrained action, analogous to the non relativistic
action. A considerable number of authors has pursued such approach 
\ci{Pavsic1, Stueckelberg, Horwitz} and
has provided the arguments why (\ref{C1}) and its quantization is good
for description of relativistic particles. In relation to the Fock-Schwinger
proper time formalism \ci{Schwinger} that was pursued by DeWitt 
\ci{DeWitt1,DeWitt2, DeWitt3}, in eq.
(\ref{C1}) we have nothing but an auxiliary, unphysical action, whose
(e.g., canonical or path integral) quantization employs the unphysical
Hilbert space and the unphysical ``evolution" parameter $\tau$
(which, in particular can be the proper time $s$). Physical (on mass shell)
states are
obtained by integrating the unphysical, $\tau$-dependent states, over $\tau$.
In path integral quantization of (\ref{C1}) one obtains unphysical,
$\tau$-dependent, Green's function from which, by integration over
$\tau$, one obtains the physical Feynman Green's functions.

In order to disentangle the problem of quantization in curved space(time) from
the intricacies of the relativistic theory whose quantization has to take
into account
the mass shell constraint, many researchers have so far chosen first to
tackle the easier problem which has roots in the unconstrained action
(\ref{C1}). The latter action was employed by DeWitt 
\ci{DeWitt1,DeWitt2, DeWitt3} in his curved-space
generalization of the Fock-Schwinger proper time technique for finding
the Feynman Green's function G(x,x').

The canonical momentum belonging to the action (\ref{C1}) is $p_\mu =
\p L/\p {\dot X}^\mu$, and its square $p_\mu p^\mu \equiv M^2$ is an
arbitrary constant of motion. The Hamiltonian is
\be
     H = {\Lambda \oo 2} g^{\mu \nu} p_\mu p_\nu = {\Lambda \oo 2} \, p^2
\lbl{C2}
\ee
where $p= \gamma^\mu p_\mu$ is the vector momentum.

Upon quatization $x^\mu$ and $p_\mu$ become operators satisfying
\be
      [x^\mu , p_\mu ] = i {\delta^\mu}_\nu
\lbl{C3}
\ee
\be
       [x^\mu, x^\nu ] = 0 \; , \qquad [p_\mu , p_\nu ] = 0
\lbl{C4}
\ee
Eigenvectors $|x' \rangle $ satisfy
\be
      x^\mu |x' \rangle = x'^\mu |x' \rangle
\lbl{C5}
\ee
and they form a complete eigenbasis.  They satisfy
the following normalization condition \ci{DeWitt1,DeWitt2}
\be
    \langle x|x' \rangle = {{\delta (x - x')}\oo {\sqrt{|g(x)|}}} =
    {{\delta(x-x')}\oo {\ggg}} \equiv \delta(x,x')
\lbl{24}
\ee
In this representation the position operator is diagonal, but the momentum
operator is not diagonal. We shall now determine its matrix elements.

Multiplying from the left by basis vectors $\gamma^\nu$ and summing over $\nu$
we obtain the commutator which involves the vector momentum $p$:
\be
    [x^\mu,p] = i \gamma^\mu
\lbl{29}
\ee

If we sandwich the commutation relations (\ref{29}) between the position 
eigenstates we have
\be
    \langle x|[x^\mu, \gamma^\nu (x) p_\nu ]|x' \rangle = (x^\mu - x'^\mu)
    \langle x|\gamma^\nu p_\nu |x' \rangle = \langle x |i \gamma^\mu|x' \rangle
    = i \gamma^\mu (x) \delta (x,x')
\lbl{29a}
\ee
From
\bear
    \gamma^\alpha (x) \p_\alpha \Bigl( (x^\mu \delta (x,x') \Bigr) 
    &=& \gamma^\alpha (x)
    \Bigl( {\delta^\mu}_\alpha \delta (x,x') + x^\mu \p_\alpha
    \delta (x,x') \Bigr) \nonumber \\
    \gamma^\alpha (x) \p_\alpha \Bigl( x'^\mu \delta (x,x') \Bigr) 
    &=& \gamma^\alpha (x) x'^\mu \p_\alpha \delta (x,x')
\lbl{30}
\ear
we obtain
\be
    (x^\mu - x'^\mu) \gamma^\alpha (x) \p_\alpha \delta (x,x') =
    - \gamma^\mu \delta (x,x')
\lbl{30a}
\ee
Comparing eqs.(\ref{29a}) and (\ref{31}) we obtain
\be
    \langle x|\gamma^\nu p_\nu|x' \rangle = -i \gamma^\nu \p_\nu \delta (x,x')
    + F(x) \delta (x,x')
\lbl{31}
\ee
where $F(x)$ is an arbitrary position dependent vector.

The same relation (\ref{31}) can be derived from the commutation relations
(\ref{C3}) which give \ci{DeWitt1,DeWitt2}
\be
    \langle x|p_\mu |x' \rangle = - i \p_\mu \delta (x,x') + F_\mu (x)
    \delta (x,x')
\lbl{32}
\ee
and contracting eq. (\ref{32}) by $\gamma^\mu$.

In the literature so far the authors considered the matrix elements (\ref{32})
and required that they had to satisfy the condition for Hermiticity
\be
    \langle x' |p_\mu |x \rangle^* = \langle x |p_\mu |x' \rangle
\lbl{33}
\ee
This together with the commutation relations $[p_\mu , p_\nu ] = 0$
restricts the choice of $F_\mu (x)$:
\be
    F_\mu = \p_\mu \left ( - {i\oo 4} \, {\rm ln} |g| - \chi \right )
\lbl{34}
\ee
where $\chi(x)$ is an arbitrary function -- a phase. Hence, for the choice
$\chi = 0$,
\be
    \langle x |p_\mu |x' \rangle = - i \left ( \p_\mu + \mbox{${1\oo 4}$} 
    \p_\mu {\rm ln} \, |g| \right ) \delta (x,x')
\lbl{35}
\ee
where 
\be
    \mbox{${1\oo 4}$} \p_\mu {\rm ln} \, |g| = \mbox{${1\oo 4}$} |g|^{-1} \p_\mu 
    |g| = \mbox{${1\oo 2}$} |g|^{-1/2} \p_\mu |g|^{1/2} =
    |g|^{-1/4} \p_\mu |g|^{1/4} = \mbox{${1\oo 2}$} \Gamma_{\mu \alpha}^\alpha
\lbl{36}
\ee

By employing the powerful geometric calculus we recognize that the
momentum operator should be defined as the vector operator
\be
    p = \gamma^\mu p_\mu
\lbl{37}
\ee
So instead of the matrix elements $\langle x |p_\mu |x' \rangle$ 
(eq.(\ref{32})) we have to consider the matrix elements $\langle x|p |
x' \rangle$ (eq.(\ref{31})). Hermiticity condition for the latter
matrix elements determines $F(x)$ in eq.(\ref{31}). It turns out (see
eqs.(\ref{26a})--(\ref{27})) that
\be
    F(x) = 0
\lbl{38}
\ee
Hence
\be
     \langle x |p |x' \rangle = - i \gamma^\mu (x) \p_\mu \delta (x,x')
\lbl{25}
\ee
and we have
\be
    \langle x |p |\phi \rangle = \int \langle x|p|x' \rangle \ggg \, \dd x' \,
    \langle x' |\phi \rangle = - i \gamma^\mu \p_\mu \phi \equiv p \, \phi
\lbl{26}
\ee
where $|\phi \rangle $ is a state vector and $\langle x | \phi \rangle \equiv
\phi$ a wave function.

We shall now demonstrate that the matrix elements (\ref{25}) are Hermitian.
Using
\be
    \p_\mu \delta (x,x') = - \p'_\mu \delta (x,x') - {1\oo \ggg} \,
    \p'_\mu \ggg \, \delta (x,x')
\lbl{26a}
\ee
which follows directly from eq.(\ref{24}), and by taking into account
eq.(\ref{20}) (see next section) and the relation
\be
      \Gamma_{\mu \alpha}^\alpha = {1\oo {\sqrt{|g|}}} \, \p_\mu \sqrt{|g|}
\lbl{27b}
\ee
we have
\be
    - i \gamma^\mu (x) \p_\mu \delta (x,x') = i \gamma^\mu (x') \p'_\mu
    \delta (x,x')
\lbl{27a}
\ee
From the latter relation it is straightforward to verify that
\be
    \langle x'|p|x |\rangle^* = \langle x|p|x' \rangle
\lbl{27}
\ee
Thus {\it the matrix elements} (\ref{25}) {\it of the vector momentum operator
in curved space satisfy the Hermiticity condition.} No extra terms to 
eq.(\ref{25}) are necessary in order to assure Hermiticity. This is not the
case in the approaches which work with components $p_\mu$ of momentum
operator: the matrix elements $\langle x|p_\mu |x' \rangle = - i \p_\mu
\delta (x,x')$ are not Hermitian.

Let us now also calculate the matrix elements of the square of the vector
momentum operator. Using eqs.(\ref{25}),(\ref{27a}) and (\ref{22}),
namely $\p_\mu (\sqrt{|g|} \gamma^\mu ) = 0$, we find
\bear
    \langle x |p^2|\phi \rangle &=& \int \langle x|p|x' \rangle \, 
    \ggg \dd^n x' \, \langle x'|p|x'' \rangle \, \sqrt{|g(x'')|}
    \dd^n x'' \, \langle x'' |\phi \rangle \nonumber \\
    &=& \int i \gamma^\mu (x') \p'_\mu \left ( {{\delta (x-x')}\oo \ggg }
    \right ) \, \ggg \dd^n x' \, \nonumber \\
    && \hs{2cm} \times \, i \gamma^\nu (x'') \p''_\nu \left (
    {{\delta (x'-x'')}\oo \sqrt{|g(x'')|}} \right ) \, \sqrt{|g(x'')|}
    \dd^n x'' \, \phi (\tau, x'') \nonumber \\
    &=& i \int \gamma^\mu (x') \p'_\mu  \left ( {{\delta (x-x')}\oo \ggg }
    \right ) \, \ggg \dd^n x' \nonumber \\
    && \hs{2cm} \times \, (-i)
    {{\delta (x'-x'')}\oo \sqrt{|g(x'')|}} \,  {\p''}_\nu \left (
    \gamma^\nu (x'') \sqrt{|g(x'')|}\, \phi(\tau, x'') \right ) \, \dd^n x''
    \nonumber \\
    &=& \int \gamma^\mu (x') \p'_\mu  \left ( {{\delta (x-x')}\oo \ggg }
    \right ) \, \ggg \dd^n x' \, \gamma^\nu (x') \p'_\nu \phi (\tau, x')
    \nonumber \\
    \nonumber \\
    &=& - \gamma^\mu \p_\mu (\gamma^\nu \p_\nu \phi) = - \DD_\mu \DD^\mu \phi
    \lbl{39}
\ear
where $\phi(\tau, x)$ is the projection of a state $|\phi \rangle $ onto
position eigenstates.
The above result explicitly demonstrates the
consistency of our procedure within the matrix formalism of quantum mechanics
in curved space. We see that the matrix element $\langle x|p^2|\phi \rangle$
can be calculated by inserting twice the complete set of the position
eigenstates so that under the integration we have the product of two terms
such as $\langle x|p|x' \rangle$. The result is the product of two vector
differential operators $-i \gamma^\mu \p_\mu$, acting on the wave
function, which, according 
to eqs.(\ref{6}--\ref{9a}), is equal to the covariant d'Alambert operator 
(multiplied by $-1$).

The quantization procedure introduced in this section does not use at all
the vector integral which was discussed in previous section. Because of the
presence of the delta function, the integrals such as (\ref{26}) and (\ref{39})
are local. The former integral gives a vector, whilst the latter integral
gives a scalar at a point $x$. The delicate vector integral will be
used in next section only in the calculation of the expectation value
for the vector momentum operator.

\section{On the Schr\" odinger representation for the momentum vector
differential operator and its expectation value}

We have seen (see eqs. (\ref{37})--(\ref{26}))
that in the Schr\" odinger (coordinate) representation
the {\it momentum operator}, that remains Hermitian in curved space, 
is a vector differential operator\footnote{
As $p_\mu = - i \p_\mu$, also the vector operator $p = \gamma^\mu p_\mu$ is
generator of translations, since $\delta \phi =$
$\p_\mu \phi \, \delta x^\mu = i p_\mu \phi \, \delta x^\mu =$
$i (p\cdot \delta x) \phi$, where $\delta x = \delta x^\mu \, \gamma_\mu $.}
\be
    p = - i \p = - i \gamma^\mu \p_\mu
\lbl{10}
\ee
acting on a wave function $\phi (\tau, x^\mu)$ which is taken here to be
a {\it scalar}. The wave function is normalized according to
\be
     \int \sqrt{|g|} \, \dd^n x \, \phi^* (\tau, x^\mu) \phi (\tau, x^\mu ) = 1
\lbl{C6}
\ee
and satisfies the Schr\" odinger equation
\be
      i {{\p \phi}\oo {\p \tau}} = H \phi
\lbl{C7}
\ee
where           
\be
    H \phi = {\Lambda \oo 2 } p^2 \phi = {\Lambda \oo 2} 
    (-i)^2 \p^2 \phi = - {\Lambda \oo 2}
    \DD_\mu \DD^\mu \phi = - {\Lambda \oo 2} \, {1\oo {\sqrt{|g|}}}
    \, \p_\mu (\sqrt{|g|} \, g^{\mu \nu} \, \p_\nu \phi)
\lbl{12}
\ee

{\it With the definition} (\ref{10}) {\it of momentum operator 
there is no ordering
ambiguity} in the expressions such as $p^2$, $p^3$, etc.\, . We have
\bear
     p \phi &=& - i \gamma^\mu \p_\mu \phi \nonumber \\
     p^2 \phi &=& (-i \gamma^\mu \p_\mu)(-i \gamma^\nu \p_\nu) \phi
        = (-i)^2 \gamma^\mu \gamma^\nu \DD_\mu \DD_\nu \phi \nonumber \\
     p^3 \phi &=&  (-i \gamma^\mu \p_\mu)(-i \gamma^\nu \p_\nu)
     (- i \gamma^\alpha
     \p_\alpha) \phi = (-i)^3 \gamma^\mu \gamma^\nu \gamma^\alpha
     \DD_\mu \DD_\nu \DD_\alpha \phi \lbl{13}\\
     {\rm etc.}&&
\ear
All those expressions are covariant with respect to arbitrary coordinate
transformations in curved space. The metric $g^{\mu \nu}$ of curved space
is implicit in the position dependent basis vectors $\gamma^\mu (x)$
satisfying eq.(\ref{1}). There is no ambiguity of where to place
$\gamma^\mu$, $\gamma^\nu$, etc., in the product of any number of operators
$p$.

Remember that we are considering here the action of the {\it geometric
derivative} $\p_\mu$ on a {\it scalar} $\phi$ so that $\p_\mu \phi$ coincides
with the partial derivative of $\phi$. Had we considered the action
of $\p_\mu$ on a spinor, the situation would be different. Consideration
of spinors is beyond the scope of this paper.

Let us now explicitly verify that the operator $p$ is self-adjoint with
respect to the scalar product
\be
   \int \dd^n x \,\sqrt{|g|} \,  \phi^* p \, \phi  = \langle p \rangle
\lbl{14}
\ee
where $\phi = \phi (\tau, x)$ depends on the evolution parameter
$\tau$ and coordinates $x^\mu$.

Eq.(\ref{14}) defines the expectation value of the momentum operator
$p$.  By using eq.(\ref{10}) we write it explicitly
\be
    \langle p \rangle = -i \int \sqrt{|g|} \,  \dd^n x \, 
    \phi^* \gamma^\mu \p_\mu
    \phi
\lbl{18}
\ee
Its complex conjugate value is\footnote{The complex conjugate vector $a^*$ 
is defined according to the relation $a^* = {a^*}^\mu \gamma_\mu$ in which
the basis vectors remain unchanged.}
\bear
    \langle p \rangle^* &=& i \int \sqrt{|g|} \,  \dd^n x \, 
    \phi \gamma^\mu \p_\mu
    \phi^* \nonumber \\
    &=& -i \int \sqrt{|g|} \,  \dd^n x \, \phi^* \gamma^\mu \p_\mu \phi -
    i \int \dd^n x \, \phi^* \p_\mu (\sqrt{|g|} \,  \gamma^\mu )
    \nonumber \\
    && \hs{1cm} + \, i \int \dd^n x \, \p_\mu (\phi^* \sqrt{|g|} \,  
    \gamma^\mu \phi)
\lbl{19}
\ear
From eq.(\ref{7a}) we have
\be
    \p_\mu \gamma^\mu = - \Gamma_{\mu \alpha}^\mu \gamma^\alpha
\lbl{20}
\ee
On the other hand, the derivative of the determinant gives
\be
    \p_\mu \sqrt{|g|} = \sqrt{|g|}\, \Gamma_{\alpha \mu}^\alpha
\lbl{21}
\ee
Therefore,
\be
     \p_\mu (\sqrt{|g|} \,  \gamma^\mu) = \p_{\mu} \sqrt{|g|} \,  \gamma^\mu +
     \sqrt{|g|} \,  \p_\mu \gamma^\mu = \sqrt{|g|} \,  
     (\Gamma_{\alpha \mu}^\alpha \gamma^\mu
     - \Gamma_{\mu \alpha}^\mu \gamma^\alpha) = 0
 \lbl{22}
 \ee
Using eq.(\ref{22}) and assuming that $\phi(\tau, x)$ vanishes at the
boundary of the integration domain $\Omega$, so that the boundary term
can be omitted, we find from (\ref{19}) that the expectation value is real:
\be
    \langle p \rangle^* = \langle p \rangle
\lbl{23}
\ee
which means that the vector momentum operator $p = - i \gamma^\mu
\p_\mu$ is self-adjoint with respect to the scalar product (\ref{14}).

The {\it boundary term} in eq.(\ref{19}) comes from the {\it boundary theorem}
which for an arbitrary polyvector $A$ in curved space $V_n$ can be written
as \ci{Hestenes}
\be
\int_\Omega  \dd \omega \cdot \p A = \int_\Sigma \dd \sigma \, A
\lbl{b1}
\ee
Here $\Omega$ is a closed $r$-dimensional volume bounded by an 
$(r-1)$-dimensional surface $\Sigma$, whilst the $r$-vector $\dd \omega$
is the volume element of $\Omega$, and the $(r-1)$-vector $\dd \sigma$ is
the surface element of $\Sigma$.

It is straightforward to show \ci{Hestenes} that in Riemanian geometry
the {\it boundary
operator} $\p$ in eq. (\ref{b1}) may be taken to coincide with the
{\it geometric derivative} defined in eqs. (\ref{5})--(\ref{7a})
and used in eqs. (\ref{18})--(\ref{23}). It cannot coincide with
the commuting partial derivative with the properties (\ref{A5})--(\ref{A7}).

The {\it expectation value} (\ref{18}), which we will now denote
$\langle p \rangle \equiv P$, is defined by the vector integral which includes
into its definition a choice of a point $x'$.
Since the wave function depends on time $\tau$,
the expectation value $\langle p \rangle = P(\tau)$ in principle also depends
on $\tau$.

Let us now calculate the derivative of the momentum expectation value
with respect to $\tau$. From
\be
    \langle p \rangle = \int \phi^* p \phi \sqrt{|g|} \, \dd^n x
\lbl{A25}
\ee
we have
\be
       {{\dd \langle p \rangle }\oo {\dd \tau}} = \int ({{\p \phi^*}\oo
       {\p \tau}} p \phi
       + \phi^* p {{\p \phi}\oo {\p \tau}}) \, \sqrt{|g|} \, \dd^n x = i \int
       ((H\phi^*) p \phi - \phi^* H \phi) \sqrt{|g|} \, \dd^n x  
\lbl{A26}
\ee
where we have taken into account the Schr\" odinger equation 
$i \p \phi/\p \tau = H \phi$ and its conjugate 
$- i \p \phi^*/\p \tau = H \phi^*$.
It is straightforward to show that for $H$ (which is Hermitian, since
$p$ is Hermitian) we obtain 
\be
    \int (H \phi^*) p \phi \, \sqrt{|g|} \, \dd^n x = 
    \int \phi^* H p \phi \, \sqrt{|g|} \, \dd^n x
\lbl{A27}
\ee
so that eq.(\ref{A26}) becomes
\be
     {{\dd \langle p \rangle }\oo {\dd \tau}} = i \int \phi^* [H,p] \phi
     \, \sqrt{|g|} \, \dd^n x  = 0
\lbl{A28}
\ee
due to the fact that $p$ commutes with $H = \Lambda p^2/2$. So we have
found that the expectation value $\langle p \rangle$ does not change with
$\tau$. 

We can also look at the $\tau$ derivative of $\langle p \rangle$
from another angle.
According to our interpretation of the vector integral as discussed before
(see eqs. (\ref{16})--(\ref{A16})), the result of the integration is
a vector at a chosen point $x'$ of the manifold. In the following we will
omit the prime and denote the chosen point by $x$.
A question arises as to which
point $x$ we should choose. A natural choice for $x$ is the expectation
value $\langle x \rangle \equiv X(\tau)$ of the particle's position. The
latter point changes with $\tau$, therefore at every $\tau$ the momentum
$P(\tau)$ is taken at different point $X(\tau)$. We can compare $P$ at
different points by means of the geometric derivative $\p_\mu$ which
performs the parallel transport of $P$ at $x^\mu + \delta x^\mu$ to the point
$x^\mu$ so that
\be
    P(x + \delta x) - P(x) \equiv \delta P = \p_\mu P \, \delta x^\mu
\lbl{BB1}
\ee
By taking $x^\mu = X^\mu (\tau)$ and $x^\mu + \delta x^\mu = 
X^\mu (\tau + \delta \tau)$ we have that $\delta x^\mu = {\dot X}^\mu \delta 
\tau$ so that
\be
     \delta P = \p_\mu P \, {\dot X}^\mu \delta \tau
\lbl{BB2}
\ee
In (\ref{BB2}) we calculated the change of $P$ along the trajectory 
$X^\mu (\tau)$ at two infinitesimally separated times $\tau$ and $\tau +
\delta \tau$. The derivative of $P$ is thus
\be
       {{\dd P}\oo {\dd \tau}} \equiv {{\dd \langle p \rangle }\oo {\dd \tau}}
        = \p_\mu \, P \, {\dot X}^\mu
\lbl{BB3}
\ee

Using (\ref{A28}) and (\ref{BB3}) we thus have
\be
     {\dot X}^\mu \p_\mu P = {\dot X}^\mu \p_\mu (P^\nu \gamma_\nu)
     = {\dot X}^\mu (\p_\mu P^\nu + \Gamma_{\mu \rho}^\nu P^\rho ) \gamma_\nu
     = \left ( {{\dd P^\nu}\oo {\dd \tau}} + \Gamma_{\mu \rho}^\nu \, 
     {\dot X}^\mu
     P^\rho \right )  \gamma_\nu = 0
\lbl{A30}
\ee           
By putting $P^\nu = (1/\Lambda) \, \dd X^\nu/\dd \tau$, where
$X^\nu \equiv \langle x^\nu \rangle$ is the expectation value of the particle's
position (i.e., the ``center" of the wave packet),
we find that (\ref{A30})
is {\it the equation of motion that follows from the classical action}
(\ref{C1}). Since, according to (\ref{A28}) the momentum vector $P$ is
a constant of motion, also its square $P^2 = P^\nu P_\nu \equiv M^2$
is constant. This gives the relation $\Lambda^{-2} {\dot X}^\mu {\dot X}_\mu
= M^2$ so that $P^\nu = M\, {\dot X}^\nu/({\dot X}^\mu {\dot X}_\mu )^{1/2}$.
Inserting the latter expression for $P^\nu$ into eq.(\ref{A30}) we have
\be
    {1\oo {\sqrt{{\dot X}^2}}} \, {\dd \oo {\dd \tau}} \, \left (
    {{{\dot X}^\nu}\oo {\sqrt{{\dot X}^2}}} \right ) +
    \Gamma_{\alpha \beta}^\nu {{{\dot X}^\alpha {\dot X}^\beta}\oo 
    {{\dot X}^2}} = 0
\lbl{22e}
\ee
which is {\it the geodesic equation}. So we have found a natural result that
{\it the expectation value of the momentum operator follows a geodetic
trajectory in our curved space}. That is, at every $\tau$, the expectation 
value $P$ is tangent to a given geodesic.

\section{Conclusion}

We have shown how the long standing problem of the ordering ambiguity
in the definition of the Hamilton operator for
a quantum point particle in curved space can be elegantly resolved
by the quantization which employs the geometric calculus based on Clifford
algebra. This is yet another besides at least two other approaches to
quantization in curved spaces, namely  ``geometric quantization"
\ci{Woodhouse} and Kleinert's path integral quantization \ci{Kleinert}, 
in which ordering ambiguities do not arise.
In most other formulatations of quantum mechanics in curved
spaces ambiguities occur due to arbitrary choice of operator ordering
prescription. Common for those formulations is that
they do not use a geometric language.
Instead, they  use the component, coordinate-based, notation of
tensor calculus, which is a very powerful mathematical tool, but
also has its limitations. Our conclusion is that momentum 
and Hamilton operator in curved  space cannot be consistently formulated
in terms of the component notation. The geometric calculus based on Clifford
algebra provides a natural definition of the momentum operator $p$ which
is Hermitian, and of the Hamilton operator which is free of the ordering 
ambiguities. Moreover, the language of geometric calculus enables us to define
and calculate the expectation value of the momentum operator which turns out 
to follow a classical geodetic line---a very reasonable result. We have also 
been able to handle consistently the matrix elements of $p$ and $p^2$ 
between the position eigenstates, by which we have further completed the 
formulation of quantum mechanics in curved space. This opens new perspective 
on the subject and its further development.

\vs{2mm}

\centerline{\bf Acknowledgement}

Work was supported by Ministry of Education, Science and
Sport of Slovenia, Grant No. PO--0517.

\end{document}